\documentclass[graphicx,11pt]{aastex}
\usepackage{color}
\usepackage{ulem}

\lefthead{}
\righthead{}

\begin{document}

\title{Meridional circulation of gas into gaps opened by giant planets in three-dimensional low-viscosity disks}

\author{\textbf{A. Morbidelli$^{(1)}$, J. Szul\'agyi $^{(1)}$, A. Crida $^{(1)}$, E. Lega $^{(1)}$, B. Bitsch $^{(1)}$, T. Tanigawa $^{(2)}$, K. Kanagawa$^{(2)}$}\\  
(1) Laboratoire Lagrange, UMR7293, Universit\'e de Nice Sophia-Antipolis,
  CNRS, Observatoire de la C\^ote d'Azur. Boulevard de l'Observatoire,
  06304 Nice Cedex 4, France. (Email: morby@oca.eu / Fax:
  +33-4-92003118) \\
(2) Institute of Low Temperature Science, Hokkaido University, Sapporo, Japan \\ 
} 

\begin{abstract}

We examine {the} gas circulation near a gap opened by a {
  giant} planet in a protoplanetary disk. We show with high resolution
{3D simulations} that the  gas flows into the gap at high
altitude over the mid-plane, at a rate dependent on viscosity. We
explain this observation with a simple conceptual model. From this
model we derive an estimate of the amount of gas flowing into a gap
opened by a planet with Hill radius comparable to the scale-height of
a layered disk {(i. e. a disk with viscous upper layer and
  inviscid midplane)}. Our estimate agrees with modern MRI simulations
(Gressel et al., 2013).  We conclude that gap opening in a layered
{disk} can not slow down significantly the runaway gas accretion
of Saturn to Jupiter-mass planets.

\end{abstract}

\section{Introduction}

Understanding what sets the terminal mass of a giant planet in a
runaway gas accretion regime is an open problem in planetary science.
Runaway gas accretion is the third stage of the classical
core-accretion scenario for the formation of giant planets (Pollack et
al., 1996). We remind the reader that in stage I, a solid core of 5-10 Earth
masses ($M_\oplus$) is formed by planetesimal accretion (or possibly
by pebble-accretion; see Lambrechts and Johansen 2012, Morbidelli and
Nesvorny, 2012). In stage II, the core starts to capture gas from the
protoplanetary disk, forming a primitive atmosphere in hydrostatic
equilibrium; the continuous accretion of planetesimals heats the
planet and prevents the atmosphere from collapsing. In stage III the
combined mass of core and atmosphere becomes large enough (the actual
mass-threshold depending on opacity and energy input due to
planetesimal bombardment) that the latter cannot remain in hydrostatic
equilibrium { anymore}; thus the planet enters in an exponential gas-accretion
mode, called runaway.

The accretion timescale in the runaway accretion mode is very fast
and, once started, can {lead to a Jupiter-mass planet} in a few
$10^4$~y (see for instance the hydro-dynamical simulations in Kley,
1999; D'Angelo et al., 2003; Klahr and Kley, 2006; Ayliffe and Bate,
2009; Tanigawa et al., 2012; Gressel et al., 2013, Szulagyi et al.,
2014) in a proto-planetary disk with mass distribution similar to that
of the Minimum Mass Solar Nebula model (Weidenschilling, 1977; Hayashi, 1981). This rapid accretion {mechanism} can explain
how giant planets form. On the other hand, there is no obvious reason for this fast  accretion to stop. As its timescale is much shorter than the proto-planetary disk life time (a few My Haisch et al., 2001), it is unlikely that the  disk disappears just in the middle of this process, raising  the question why Jupiter and Saturn and many giant extra-solar
planets did not grow beyond Jupiter-mass.

It is well known that giant planets open gaps in the protoplanetary
disk around their orbits (Lin and Papaloizou, 1986; Bryden et al.,
1999). Thus it is natural to expect that the depletion of gas in the
planet's vicinity slows down the accretion process.  Still, all of the
hydro-dynamical simulations quoted above that feature the gap-opening process
show that the mass-doubling time for a Jupiter-like planet is not
longer than $10^5$ years.

However, these simulations may have been hampered by the assumption of
a prescribed viscosity throughout the protoplanetary disk, or by
significant numerical viscosity in the simulation scheme.  It is
expected that planets form in dead zones of the protoplanetary disk
(Gammie, 1996), where the viscosity is much smaller than in numerical
simulations, so that Jovian-mass planets could presumably open much wider 
and deeper gaps, with consequent inhibition of further growth 
(e.g. Thommes et al., 2008; Matsumura et al., 2009; Ida and Lin, 2004). 

%This brings us to the object of this Note. There 
{On this issue, it is worth stressing that there is quite of a
  confusion on the role of viscosity in gap opening. In a 2D disk
  isothermal model, Crida et al. (2006) showed that the width and the
  depth of a gap saturates in the limit of vanishing viscosity.  This
  paper has been challanged recently by Duffell and MacFadyen (2013)
  and Fung et al. (2013), still for 2D disks. For a massive planet,
  the results of Duffell and MacFadyen actually agree with those of
  Crida et al., because the former group also finds that the gap has a
  saturated depth and width in the limit of null viscosity and they
  demonstrate that this result is not due to numerical viscosity. The
  actual disagreement is on the ability of small planets to open gaps
  in disks much thicker than their Hill sphere. Fung et al. address
  this case just by extrapolation of formul\ae\ obtained for massive
  planets in viscous disks, so it is not very compelling. We believe
  that gap opening by small planets in Duffel and MacFadyen is due to
  {the use of an adiabatic equation of state $P=E_{\rm int}(\gamma-1)$ which, despite adopting a value for the  
    parameter $\gamma$ very close to unity,} is
  not equivalent to the isothermal case (see Paardekooper and Mellema,
  2008). The issue, however, deserves further scrutiny.

This controversy is nevertheless quite academic, because real disks have a
3 dimensional structure. Thus, in this Note we discuss gap opening in 3D disks
and we focus on giant planets that are massive enough to undergo runaway gas accretion, i.e. Jupiter-mass bodies. In Sect.~2 we present 
the structure of the gaps and the gas circulation in their vicinity, using
  three-dimensional isothermal simulations. We also interpet the results with 
a simple intuitive
  model.  From this model}, we derive in Section 3 an estimate for
  the flow of gas into a gap in a layered disk (i.e. a disk that is
  viscous on the surface and ``dead'' near the mid-plane), that is in
  agreement with the numerical results of Gressel et al. (2013).
  Conclusions and discussion of the implications for terminal mass
  problem of giant planet are reported in Sect. 4.

\section{Gaps in 3D disks}

\begin{figure}
\includegraphics[width=\linewidth]{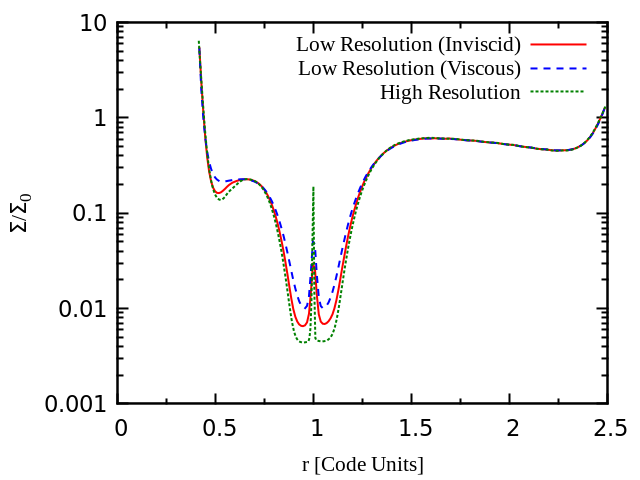}
\caption{Gap profiles in the simulations by Szulagyi et
  al. (2014). The perturbed column density distribution is normalized
  by the original column density. Here ``Viscous'' means and $\alpha$
  parameter of 0.004 in the Shakura and Sunyaev (1973) viscosity
  prescription and ``inviscid'' means $\alpha=0$; ``Low resolution''
  means 628x208x15 cells for the directions of azimuth, radius,
  co-latitude, respectively, but with mesh refinement in the vicinity
  of the planet {(we use 7 levels of mesh refinements, starting
    from a box centered on the planet with half-sizes $0.27\times 0.27
    \times 0.12$ in code units resolved in $112\times 112\times 24$
    cells, and doubling the resolution at each level)}; ``High
  resolution'' means twice the number of cells in each direction {
    at each level} and still $\alpha=0$.  Notice that the gap becomes deeper
  with decreasing viscosity and increasing resolution, i.e. decreasing
  the effective viscosity (prescribed or numeric). {The simulation has been performed with the code JUPITER written by F. Masset}.}
\label{gap3Djudit}
\end{figure}

In the framework of a study on planet accretion (Szulagyi et al.,
  2014), we conducted 3D simulations of a Jupiter-mass planet embedded
  in an isothermal disk with scale height of 5\% and
  $\alpha$-prescription of the viscosity (Shakura and Sunyaev, 1973).
  We adopted $\alpha=4\times 10^{-3}$ (viscous simulation) and
  $\alpha=0$ (inviscid simulation). The latter simulation was conducted with two
  different resolutions, to change the effective numerical viscosity
  of the simulation code.  The technical parameters of the simulations
  are described in detail in Szulagyi et al. (2014) {and the main ones are briefly reported in the caption of Fig.~\ref{gap3Djudit}.}

In Szulagyi et al. (2014) we reported that the flux of material
into the gap (and therefore the accretion rate of the planet)
decreases by a factor of 2 when the resolution is increased by a
factor of 2 (and hence the numerical viscosity is halved). {This
suggested that the gap properties do not saturate in the limit of small
viscosity (prescribed or numeric), or at least not in the range of
resolution that we have been able to attain. This observation 
motivated a deeper investigation, which is the object of the present work.}

Fig.~\ref{gap3Djudit} gives a clear demonstration of how the gap
profile opened by a Jupiter-mass planet in 3D simulations changes with
prescribed and numerical viscosity (i.e. with resolution). Each curve
shows the radial profile of the surface density, averaged over the
azimuth. The surface density has
been computed by integrating the volume density over the vertical
direction.  Clearly, here no convergence is achieved with resolution.

\begin{figure}
\includegraphics[width=\linewidth]{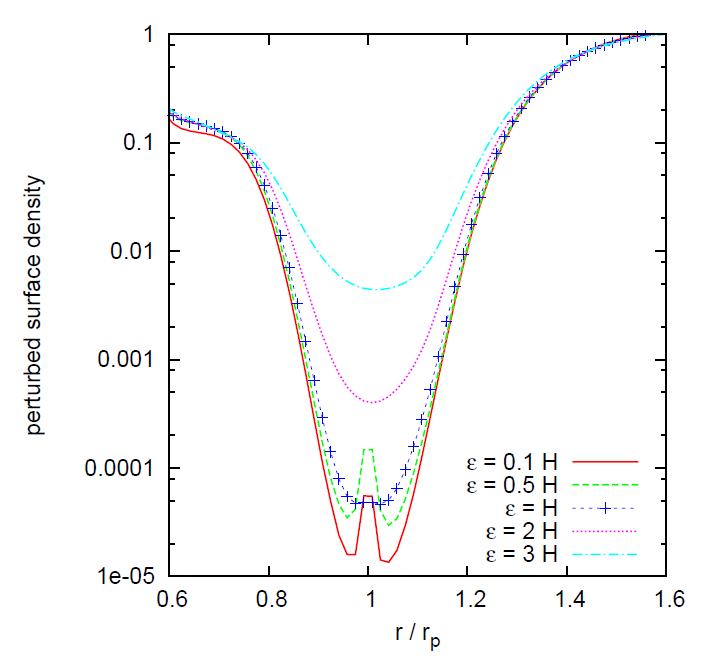}
\caption{Perturbations of the azimuthally averaged surface density
  profile in a 2D disk, after 5000 orbits in the vicinity of a Jupiter
  mass planet on fixed orbit. The size of the cells is $\delta r=0.0167$ and $\delta\theta=0.0193$. We have checked that this resolution is enough (doubling the number of cells doesn't affect the shape of the gap. The
  viscosity is $\nu=10^{-6.5} r_p^2\Omega_p$ (independent of radius). Different colors show results obrained with different 
  values of the smoothing length $\epsilon$ of the planetary
  potential. This shows the crucial role of the smoothing parameter
  which, in a 3D disk, is played by the vertical distance from the
  midplane.}
\label{fig:eps}
\end{figure}

We explain this fact with the following simple model. Consider first
the gas in the mid-plane. Its dynamics has to be similar to that in a
2D disk model, and therefore the planet opens a gap with a given
profile independent of viscosity in the small viscosity limit.  
In the vertical direction, the disk has to be in hydrostatic equilibrium.
This implies that the volume density $\rho$ scales with $z$ (the
distance from the mid-plane) as
\begin{equation}
\rho(z)=\rho(0) \exp\left(-\frac{z^2}{2H^2}\right)\ ,
\label{rho-z}
\end{equation} 
where $\rho(0)$ is the mid-plane density and $H$ is the scale-height of
the disk. Therefore, at equilibrium the radial density profile of the
gap has to be the same at all altitudes. However, the planet cannot
sustain the same profile at high altitude as on the mid-plane, because
its gravitational potential:
\begin{equation}
\Phi(d,z)={\frac{{\cal{G}} M_p}{\sqrt{d^2+z^2}}}
\end{equation}  
(where $\cal{G}$ is the gravitational constant, $M_p$ is the planet's
mass, $d$ is the planet-fluid element distance projected on the
mid-plane) weakens with increasing $z$.  Please notice that $z$ plays
the role here of the smoothing parameter (usually denoted $\epsilon$)
used in 2D simulations, and it is well known that the depth of the gap
decreases with increasing smoothing length (see
Fig.~\ref{fig:eps})\footnote{This shows that neglecting the
  $z$-dependence of the potential in the numerical simulations, as
  done in Zhu et al. (2013), changes the dynamics of the gas at a
  qualitative level and can enable small-mass planets to sustain
  partial gaps.}.

 Away from the mid-plane, the gas tends to refill the gap by viscous diffusion
because it is not sufficiently repelled by the { planetary
  torque}. As soon as the gas penetrates into the gap, however, it has
to fall towards the mid-plane because the relationship (\ref{rho-z})
is no longer satisfied and has to be restored. Therefore, there is
always more gas near the mid-plane in the gap than it would in the
ideal case where vertical motion is disabled.  This excess of gas is
partially accreted by the planet and partially repelled away from the
planet orbit into the disk, like in the gap opening process.  Outside
of the gap, then, the relationship (\ref{rho-z}) is also not
fulfilled, because there is an excess of gas near the mid-plane,
coming from within the gap, and a deficit of gas at high altitude,
which flowed into the gap. So, the gas has to move towards the surface
of the disk to restore the hydrostatic vertical equilibrium profile
(\ref{rho-z}).

{In conclusion, there has to be a 4-step meridional circulation in
the gas dynamics: (1) the gas flows into the gap at the top layer of
the disk; (2) then it falls towards the disk's mid-plane; (3) the
planet keeps the gap open by accreting part of this gas and by pushing
back into the disk the gas that flowed outside of the Hill sphere (4)
the gas expelled from the gap near the mid-plane rises back to the
disk's surface. If no gas were permanently trapped in the vicinity of
the planet (i.e. no planet growth), this meridional circulation would
basically be a closed loop. Instead, planet accretion makes the flow at step (3)
smaller than that at step (1)}

\begin{figure}
\vskip -3truecm
\includegraphics[width=13truecm,height=13truecm]{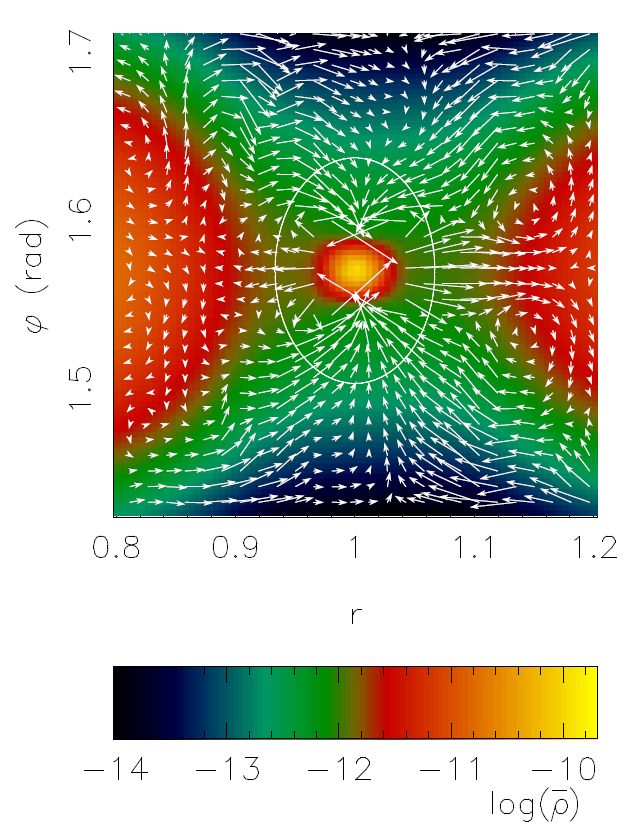}
\caption{Volume density (in $g\over cm^3$) of the disk plotted in
  $r,\phi$ coordinates, integrated over the azimuth. The arrows show
  the mass transport $\rho \vec{v}$, also averaged over the azimuth.
  This simulation has been done with the 3D version of FARGO developed
  in Lega et al. (2013). The viscosity here is $\nu=10^{-5} r_p^2\Omega_p$
  (independent of radius) and the resolution is $760,1152,96$ in
  radius, azimuth and co-latitude respectively, with radius ranging
  from $r_{min}=0.3$ to $r_{max}=4.2$ (in units of the orbital radius
  of the planet) and co-latitude from $1.42$ to
  $1.72$. The Jupiter mass planet is held fixed at $r_p=1$,
  $\phi = \pi/2$ (midplane). {The planet is not allowed to accrete (no mass sink is implemented) so that, by viscosity, most of the gas that falls onto the circum-planetary disk eventually flows outside of the planet's Hill sphere. Consequently, the meridional circulation shown in the figure has practically reached a steady state as a closed loop.}}
\label{densvel}
\end{figure}

We can observe this meridional loop in the numerical simulations.
Fig.\ref{densvel} shows the volume density of the disk in $r,\phi$
coordinates and the arrows show the mass
transport, both
averaged over the azimuth. Kley et al. (2001), Ayliffe and Bate
(2009), Machida et al. (2010), Tanigawa et al. (2012), Gressel et
al. (2013) and Szulagyi et al. (2014) already showed that gas flows
into the gap near the surface of the disk, but Fig.\ref{densvel} is
the first clear portrait of the meridional circulation explained
above.

Notice that steps 2 to 4 of the meridional circulation occur on a
dynamical timescale (i.e. independent of viscosity), because they are
related to pressure and planet's gravity. Step 1, instead, occurs on a
viscous timescale. If the viscosity is small, the viscous timescale
controls the entire timescale of the meridional circulation.  This
explains why the flow into the gap and the accretion rate onto the
planet scale linearly with effective viscosity (prescribed or numeric)
as found in Szulagyi et al. (2014). Moreover, because step 1 and step
3 occur on independent timescales, the depth of a gap also increases
with decreasing viscosity, as observed. The relationship, though, is
not necessarily linear. Part of the gas falling toward the mid-plane
in the gap ends in the horseshoe libration region and has to diffuse
to the separatrices of said region (on a viscous timescale) before
being removed. If all gas followed this fate, the inflow and the
removal would both happen on the same viscous timescale and therefore
the gap's depth would not change with viscosity. If instead all the
gas fell on the separatrices of the horseshoe region, only the inflow
would depend on viscosity and therefore the gap's depth would be
inversely proportional to the viscosity. Reality is in between the two.  In
fact, in Fig.~\ref{gap3Djudit}, the gap's depth increases by a factor
$\sim 1.5$ as the resolution is doubled (e.g. the numerical viscosity
is halved).

\section{Implications on planet's accretion from the flow of gas into the gap }

The results of the previous section seem to suggest that the flow of
gas into the gap and the planet's accretion rate have to vanish with
viscosity. But remember that what governs the meridional circulation of
gas in the gap's vicinity is the viscous timescale near the disk's
{\it surface}. Protoplanetary disks cannot be fully inviscid: stars
are observed to accrete mass, which implies that angular momentum has
to be transported at least in part of the disk. A popular view is that
of a layered disk, i.e. a disk which is viscous near the surface and
inviscid near the mid-plane, due to ionization at high altitude and
turbulence driven by the magneto rotational instability (MRI; Gammie,
1996). This view may have problems (see Turner et al. 2013 for
a review), so that other mechanisms (e.g. the baroclynic instability;
Klahr, 2004) may be at work. Possibly these alternative mechanism can
make the entire protoplanetary disk viscous, but at the very least, a
viscous layer has to exist near the surface of the disk, as in the MRI
view.

We can now make a simple estimate of the gas-flow into a gap, as follows.
It is well known that the radial velocity of the gas in a {steady state} viscous disk is:
\begin{equation}
v_r= \frac{1}{\Sigma j}\frac{d}{dr}[j \nu\Sigma]
\label{vr}
\end{equation}
where $\nu$ is the viscosity, $j=\sqrt{GM_\odot r}$ is the specific
angular momentum and $\Sigma$ is the surface density of the disk,
which is not a constant slope any more, because the profile changed
due to the opened gap. From now on we set the units so that the mass
of the star $M_\odot$ and the gravitational constant $G$ are both
equal to 1, so that we substitute $j$ with $\sqrt{r}$.

Assuming that $\nu=\alpha H^2 \Omega$ as in the usual
$\alpha$-model (Shakura and Sunyaev, 1973), where $\Omega$ is the
orbital frequency of the gas, one has: $\nu=\alpha (H/r)^2 j$.
Thus, the right-hand side of eq. (\ref{vr}) becomes the sum of two
terms. Assuming that $H/r$ is independent of $r$, { we get
\begin{equation}
 v_r = \alpha \left(\frac{H}{r}\right)^2
\left(1+\frac{d\ln\Sigma}{d\ln r}\right)r\Omega\ .
\label{vr2}
\end{equation}
}

According to Crida et al. (2006) (see formula 14 in that paper, with input from formul\ae\ 11 and 13) in the
limit of vanishing viscosity and for a planet with Hill radius
$R_h\sim H$, the surface density slope
is maximum at a distance $\Delta=2.5R_h$ and is:
\begin{equation}
\frac{d(\log\Sigma)}{dr}= \frac{1.3}{r_p(H/r)}\ ,
\label{slope}
\end{equation}
where $r_p$ is the orbital radius of the planet.
As we explained in the previous section, the same slope is present at
every altitude because of the {hydrostatic equilibrium} relationship in (\ref{rho-z}). At high
altitude, therefore, this is the slope that drives the viscous flow,
given that the planet can not sustain the gap there, due to its
reduced potential. From  (\ref{slope}) the second term of the right-hand side of
(\ref{vr2}) is inversely proportional to $H/r$ and consequently it dominates over the
first term. Thus, we drop the first term from the formula, getting
\begin{equation}
v_r\sim \alpha (H/r) r\Omega\ .
\label{newvr}
\end{equation}

The flow in the gap is then
\begin{equation}
\dot{M}=2\pi r v_r 4\Sigma_a
\label{flux}
\end{equation}
where $\Sigma_a$ is the column density of the active layer of the
disk, typically 10g/cm$^2$ in the MRI case, if ionization is due to
X-ray penetrating into the disk (Igea and Glassgold, 1999). Notice
that factor of 4 multiplying $\Sigma_a$, due to the fact that there
are two surface layers in a disk and two sides of the gap.

Let's now take formul\ae\ (\ref{flux}) and (\ref{newvr}) and apply
them to a planet with $R_h\sim H$ at 3.5 AU. Assuming $\alpha =
3\times 10^{-3}$, a typical value for the active layer of a MRI disk
(Gressel et al., 2011) and $H/r=0.05$, formula (\ref{newvr}) gives
$v_r= 7.9\times 10^{9}$ cm/y and $\dot{M}=10^{26}$g/y, i.e.
$1.7\times 10^{-2} M_\oplus$/y. If half of this flux is accreted by the central planet, this value matches the accretion rate
of a Saturn-mass planet observed in the MRI simulations of Gressel et
al. (2011).

\section{Conclusions}

In this paper we have analyzed the dynamics of gas near gaps{,} opened by
planets in three dimensional proto-planetary disks, with particular emphasis on the 
small viscosity limit.

We observe a flow of gas into the gap at high altitude in the
disk, at a rate dependent on effective viscosity (prescribed or
numeric). We have explained this result with a simple analytic model,
that assumes that the disk is in vertical hydrostatic equilibrium at
all radii. Thus the radial profile of the gas volume density is the
same at every altitude in the disk. However, the planet's potential
weakens with altitude $z$, so that the planet can not sustain the same
gap on the mid-plane and at the surface of the disk. Consequently, gas
flows into the gap, at a viscous rate, at high altitude. Only planets
with Hill radius $R_h >> H$ can sustain the same gap at all altitudes
$z<H$, but this occurs only for very massive planets (typically many
Jupiter masses).

Assuming that proto-planetary disks are layered, as in the MRI
description, we have derived from our model a simple formula for the
amount of gas flowing into a gap opened by a planet with $R_h\sim H$,
which is in good agreement with the MRI simulations of Gressel et al
(2013).  The mass flux is large, corresponding to a doubling time for
the mass of Jupiter of about 50,000~y.

We conclude that gap opening can not be the answer to the terminal
mass problem of giant planets, described in the introduction. A
different mechanism is needed in order to slow down the gas accretion
of planets of Saturn to Jupiter mass. The role of the circum-planetary
disk is a promising one, as shown in Rivier et al. (2012) and Szulagyi
et al. (2014), although this needs to be explored further with more
realistic simulations.

\acknowledgments

The Nice group is thankful to ANR for supporting the MOJO project. {J. Szul\'agyi acknowledges the support from the Capital Fund
Management's J.P. Aguilar Grant.} {The computations have been done on the ``Mesocentre SIGAMM" machine, hosted 
by the Observatoire de la C\^ote d'Azur.}  
T.T. is supported by Grant-in-Aid
for Scientific Research (23740326 and 24103503) from the MEXT of
Japan. We also thank P. Duffell for open and frank discussions on gap opening by small planets in 2D disks.

%\begin{thebibliography}{}

\section{References}

\begin{itemize}

\item[--]  Ayliffe, B.~A., Bate, 
M.~R.\ 2009.\ Gas accretion on to planetary cores: three-dimensional 
self-gravitating radiation hydrodynamical calculations.\ Monthly Notices of 
the Royal Astronomical Society 393, 49-64. 
\item[--] Bryden, G., Chen, X., 
Lin, D.~N.~C., Nelson, R.~P., Papaloizou, J.~C.~B.\ 1999.\ Tidally Induced 
Gap Formation in Protostellar Disks: Gap Clearing and Suppression of 
Protoplanetary Growth.\ The Astrophysical Journal 514, 344-367. 
\item[--]  Crida, A., Morbidelli, 
A., Masset, F.\ 2006.\ On the width and shape of gaps in protoplanetary 
disks.\ Icarus 181, 587-604. 
\item[--] D'Angelo, G., Henning, 
T., Kley, W.\ 2003.\ Thermohydrodynamics of Circumstellar Disks with 
High-Mass Planets.\ The Astrophysical Journal 599, 548-576. 
\item[--] Duffell, P.~C., 
MacFadyen, A.~I.\ 2013.\ Gap Opening by Extremely Low-mass Planets in a 
Viscous Disk.\ The Astrophysical Journal 769, 41. 
\item[--] Fung, J., Shi, J.-M., 
Chiang, E.\ 2013.\ How Empty are Disk Gaps Opened by Giant Planets?.\ ArXiv 
e-prints arXiv:1310.0156. 
\item[--] Gammie, C.~F.\ 1996.\ Layered 
Accretion in T Tauri Disks.\ The Astrophysical Journal 457, 355. 
\item[--]  Gressel, O., Nelson, 
R.~P., Turner, N.~J.\ 2011.\ On the dynamics of planetesimals embedded in 
turbulent protoplanetary discs with dead zones.\ Monthly Notices of the 
Royal Astronomical Society 415, 3291-3307. 
\item[--] Gressel, O., Nelson, 
R.~P., Turner, N.~J., Ziegler, U.\ 2013.\ Global hydromagnetic simulations 
of a planet embedded in a dead zone: gap opening, gas accretion and 
formation of a protoplanetary jet.\ ArXiv e-prints arXiv:1309.2871. 
\item[--] Haisch, K.~E., Jr., 
Lada, E.~A., Lada, C.~J.\ 2001.\ Disk Frequencies and Lifetimes in Young 
Clusters.\ The Astrophysical Journal 553, L153-L156. 
\item[--] Hayashi, C.\ 1981. Structure of the Solar
Nebula, Growth and Decay of Magnetic Fields and Effects of Magnetic
and Turbulent Viscosities on the Nebula.\ {\it Progress of Theoretical
Physics Supplement} 70, 35-53.
\item[--] Ida, S., Lin, D.~N.~C.\ 
2004.\ Toward a Deterministic Model of Planetary Formation. II. The 
Formation and Retention of Gas Giant Planets around Stars with a Range of 
Metallicities.\ The Astrophysical Journal 616, 567-572. 
\item[--] Igea, J., 
Glassgold, A.~E.\ 1999.\ X-Ray Ionization of the Disks of Young Stellar 
Objects.\ The Astrophysical Journal 518, 848-858. 
\item[--] Klahr, H.\ 2004.\ The Global 
Baroclinic Instability in Accretion Disks. II. Local Linear Analysis.\ The 
Astrophysical Journal 606, 1070-1082. 
\item[--] Klahr, H., Kley, W.\ 2006.\ 3D-radiation hydro simulations of disk-planet interactions. I. Numerical algorithm and test cases.\ Astronomy and Astrophysics 445, 747-758. 
\item[--] Kley, W.\ 1999.\ Mass flow and 
accretion through gaps in accretion discs.\ Monthly Notices of the Royal 
Astronomical Society 303, 696-710. 
\item[--] Kley, W., D'Angelo, G., 
Henning, T.\ 2001.\ Three-dimensional Simulations of a Planet Embedded in a 
Protoplanetary Disk.\ The Astrophysical Journal 547, 457-464. 
\item[--] Lambrechts, M., Johansen, A.\ 2012.\ Rapid growth of gas-giant cores by pebble accretion.\ Astronomy and Astrophysics 544, A32. 
\item[--] Lega, E., Crida, A., Bitsch, B. and Morbidelli, A. 2013. Migration of Earth-size planets in 3D radiative
discs. Submitted.
\item[--] Lin, D.~N.~C., 
Papaloizou, J.\ 1986.\ On the tidal interaction between protoplanets and 
the primordial solar nebula. II - Self-consistent nonlinear interaction.\ 
The Astrophysical Journal 307, 395-409.
\item[--] Machida, M.~N., Kokubo, 
E., Inutsuka, S.-I., Matsumoto, T.\ 2010.\ Gas accretion onto a protoplanet 
and formation of a gas giant planet.\ Monthly Notices of the Royal 
Astronomical Society 405, 1227-1243.  
\item[--]  Matsumura, S., 
Pudritz, R.~E., Thommes, E.~W.\ 2009.\ The Growth and Migration of Jovian 
Planets in Evolving Protostellar Disks with Dead Zones.\ The Astrophysical 
Journal 691, 1764-1779. 
\item[--]  Morbidelli, A., Nesvorny, D.\ 2012.\ Dynamics of pebbles in the vicinity of a growing planetary embryo: hydro-dynamical simulations.\ Astronomy and Astrophysics 546, A18. 
\item[--] Paardekooper, S.-J., Mellema, G.\ 2008.\ Growing and moving low-mass planets in non-isothermal disks.\ Astronomy and Astrophysics 478, 245-266. 
\item[--] Pollack, J.~B., 
Hubickyj, O., Bodenheimer, P., Lissauer, J.~J., Podolak, M., Greenzweig, 
Y.\ 1996.\ Formation of the Giant Planets by Concurrent Accretion of Solids 
and Gas.\ Icarus 124, 62-85. 
\item[--] Rivier, G., Crida, A., Morbidelli, A., Brouet, Y.\ 2012.\ Circum-planetary discs as bottlenecks for gas accretion onto giant planets.\ Astronomy and Astrophysics 548, A116. 
\item[--] Shakura, N.~I., 
Sunyaev, R.~A.\ 1973.\ Black holes in binary systems. Observational 
appearance..\ Astronomy and Astrophysics 24, 337-355. 
\item[--] Szul\'agyi, J., Morbidelli, A., Crida, A. and Masset, F. 2014. 
Accretion of Jupiter-mass Planets in the Limit of Vanishing Viscosity. ApJ in press; . arXiv:1312.6302.
\item[--] Tanigawa, T., Ohtsuki, 
K., Machida, M.~N.\ 2012.\ Distribution of Accreting Gas and Angular 
Momentum onto Circumplanetary Disks.\ The Astrophysical Journal 747, 47. 
\item[--] Thommes, E.~W., 
Matsumura, S., Rasio, F.~A.\ 2008.\ Gas Disks to Gas Giants: Simulating the 
Birth of Planetary Systems.\ Science 321, 814. 
\item[--] Turner, N.~J., Lee, 
M.~H., Sano, T.\ 2013.\ Magnetic Coupling in the Disks Around Young Gas 
Giant Planets.\ ArXiv e-prints arXiv:1306.2276. 
\item[--] Weidenschilling, S.~J.\ 1977.\ The distribution of mass in the planetary system and solar nebula.\ Astrophysics and Space Science 51, 153-158. 
\item[--] Zhu, Z., Stone, J.~M., 
Rafikov, R.~R.\ 2013.\ Low-mass Planets in Protoplanetary Disks with Net 
Vertical Magnetic Fields: The Planetary Wake and Gap Opening.\ The 
Astrophysical Journal 768, 143. 

\end{itemize}

%\end{thebibliography}
\end{document}